# STRONG RESONANCES AT HIGH EXCITATION ENERGY IN $^{17}$O+ALPHA RESONANCE SCATTERING

© 2019 D. K. Nauruzbayev[1,2], A. K. Nurmukhanbetova[3], V. Z. Goldberg[4], M. La Cognata[5], A. Di Pietro[5], P. Figuera[5], S. Cherubini[5], M. Gulino[5], L. Lamia[5], R. G. Pizzone[5], R. Spartà[5], A. Tumino[5], A. Serikov[6,7] and E. M. Gazeeva[6,7]

[1]*National Laboratory Astana, Nazarbayev University, Nur-Sultan, 010000 Kazakhstan*

[2]*Saint Petersburg State University, Saint Petersburg, Russia*

[3]*Nazarbayev University, Nur-Sultan, Kazakhstan*

[4]*Cyclotron Institute, Texas A&M University, College station, Texas, USA*

[5]*Istituto Nazionale di Fisica Nucleare, Catania, Italy*

[6]*Joint Institute for Nuclear Research, Dubna, Russia*

[7]*Dubna State University, Dubna, Russia*

The Thick Target Inverse Kinematic (TTIK) approach was used to measure excitation functions for the elastic $^{17}$O ($\alpha, \alpha$) scattering at the initial $^{17}$O beam energy of 54.4 MeV. We observed strong peaks corresponding to highly excited $\alpha$-cluster states in the $^{21}$Ne excitation energy region of 8-16 MeV, which have never been investigated before. Additional tests were done at a $^{17}$O beam energy of 56.4 MeV to estimate a possible contribution of resonance inelastic scattering.

## INTRODUCTION

$^{20}$Ne nucleus is a famous textbook example of $\alpha$-clustering. On the other hand, practically nothing is known on $\alpha$-cluster structure in $^{21}$Ne. The low neutron binding energy in $^{21}$Ne makes this case much more special than, for instance, nearby $^{19}$Ne. The properties (in particular, widths) of $\alpha$-cluster states in $^{21}$Ne should be sensitive to single particle admixture. The $^{20}$Ne levels can undergo neutron decay only at excitation energy above 17 MeV, it is not surprising that the states below this energy decay by $\alpha$ particles. On the contrary, the threshold for neutron decay is the lowest in $^{21}$Ne (6.7 MeV).

A combination of $^{20}$Ne collective deformation with the extra neutron, and with the $\alpha$-cluster degree of freedom can be also an interesting issue.

There were no data on $^{21}$Ne because of the experimental difficulties of dealing with a low abundance gas target and a detection of low energy $\alpha$ particles in backward hemisphere, where the

---

[1] dosbol.nauruzbayev@nu.edu.kz

Rutherford scattering is minimal. In addition, an R-matrix analysis of resonance scattering of α particle on $^{17}$O with spin 5/2$^+$ is difficult.

Our research group measured [1] the $^{17}$O+α elastic scattering excitation function, up to a $^{21}$Ne excitation energy of 11 MeV, at the Nur-Sultan cyclotron. Very strong peaks were observed close to the high energy limit of the measurements, 11 MeV excitation energy [1]. However, the R-matrix analysis was performed only up to the $^{21}$Ne excitation energy of 9.5 MeV. We didn't complete the interpretation of the high energy data [1] due to the unknown influence of possible resonances at the higher excitation energy and to the limit of our measurements (DC-60 cyclotron [2] can provide heavy ion beams with energies up to 1.9 MeV/A).

The measurement of the $^{17}$O (α,α) resonance elastic scattering at a beam energy of 3.5 MeV/A was done at the INFN-LNS Tandem [3] by using the TTIK method [4, 5].

The scattering chamber with a diameter of 200 cm was filled up with $^4$He gas (purity 99.9999%) which served as the target gas. We used a Havar foil of 2.5 µm thickness for separation beam line from the scattering chamber. The pressure in the chamber was 149 torr so that the beam stopped in the target gas at a distance of 79.8 cm from a zero degrees detector. A thin Au foil (200 µg/cm$^2$) was placed at 147 mm from the Havar foil downstream of the beam to provide for the Rutherford scattering detected by monitor detectors.

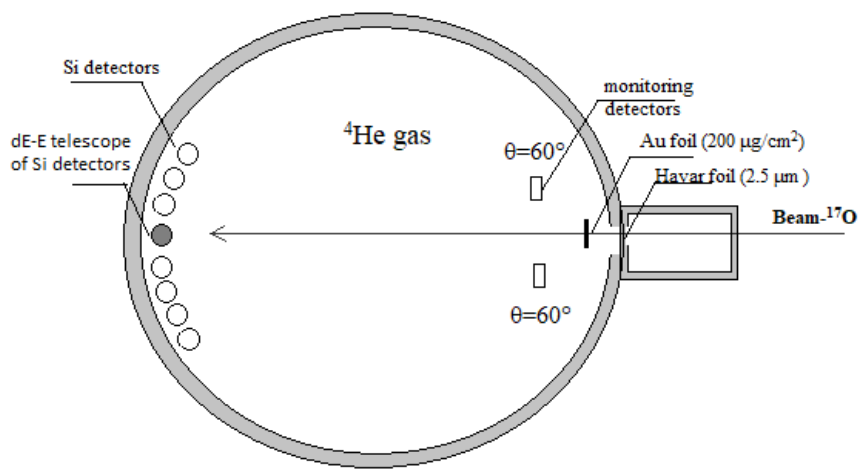

Fig. 1 the scheme of a scattering chamber for the TTIK method used at Tandem-INFN

The experimental setup is illustrated in Fig.1. An array of single Si detectors of 500 μm thickness and one dE-E telescope of Si detectors (75 and 1080 μm) were placed in the forward hemisphere to detect light recoils at different angles including 0° (laboratory system) in steps of 5°.

The dE-E telescope was needed to evaluate the contribution of protons to charge particle spectra. Additionally, we installed two Si detectors at 60°(laboratory system) with respect to Au foil and monitored beam intensity by measuring elastic scattering from it.

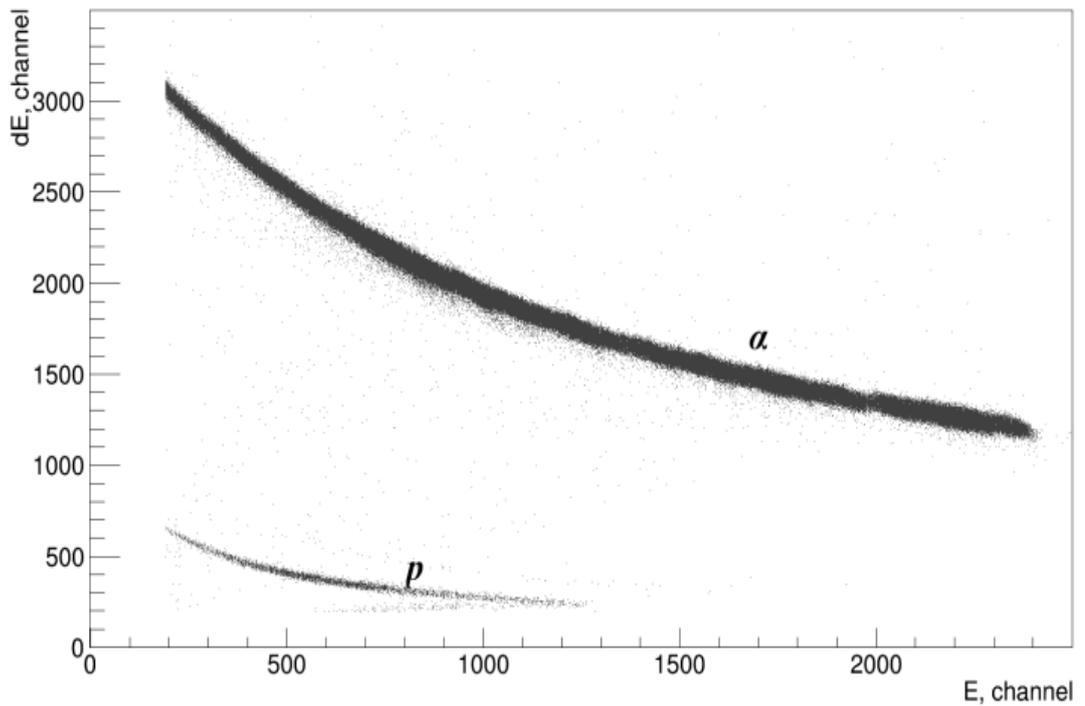

Fig. 2 dE-E spectrum of $^{17}$O+$^{4}$He interaction.

Fig. 2 demonstrates dE-E particle identification. The locus at the top in Fig. 2 corresponds to the $^{17}$O($\alpha,\alpha$) scattering. Evidently, the proton contribution to the single energy spectra is small.

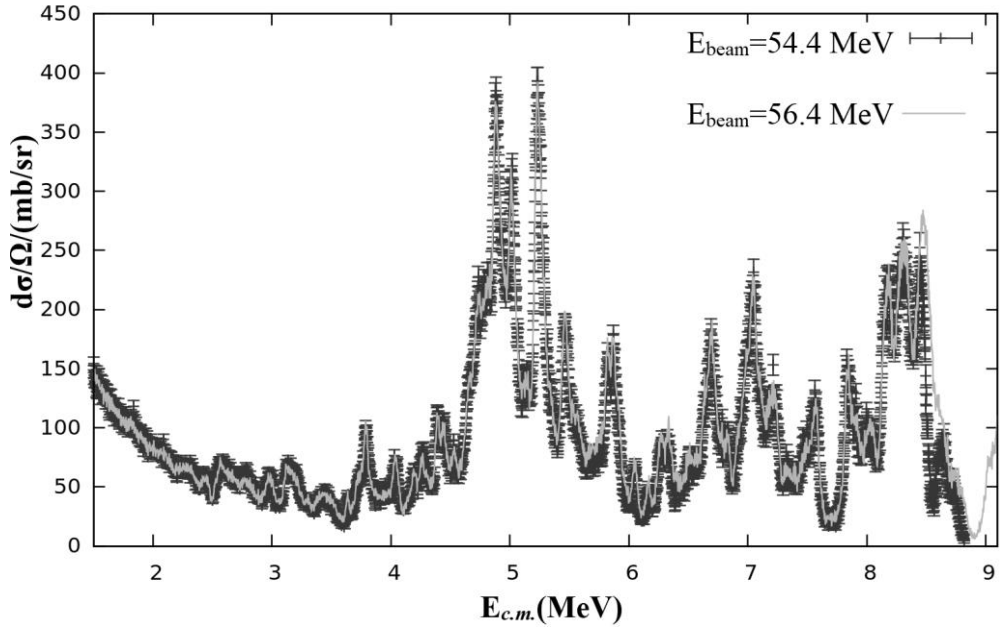

Fig.3 the excitation function of $^{17}$O+α elastic scattering with $^{17}$O beam energy of 54.4 MeV (bold line) and 56.4 MeV (solid line).

Fig. 3 shows the results of the new measurements of the excitation functions of $^{17}$O (α,α)$^{17}$O elastic scattering at 180°degree in the *c.m.* system with a higher beam energy. As seen in Fig.3, strong groups of α-cluster states manifest themselves up to the energy of 8.6 MeV. The new data will be very useful for a reliable interpretation of the observed structure in Ref. [1]. We are working on an R-matrix of the new data.

Using a simple "only E" approach in TTIK method, one cannot separate elastic and inelastic scattering. While the dominant peaks should be related with elastic scattering, and an inelastic scattering resonance cannot appear without the corresponding elastic resonance, still an experimental identification of the inelastic scattering events is desirable. To make this we made additional measurements at conditions slightly changed. We changed the beam energy by 3.6 % and kept the same gas pressure in the scattering chamber. A peak corresponding to the inelastic scattering with excitation of 3.05 MeV state in $^{17}$O at 56.4 energy beam of MeV $^{17}$O should shift relative to elastic scattering peaks by ~30 keV at the new conditions. The elastic and inelastic scattering can produce α particles of the same energy only if the inelastic scattering takes place at a higher excitation energy of the $^{17}$O beam, closer to the entrance window. However, the energy loss of $α$ particles resulted from the inelastic scattering increases relative to that of the elastic scattering. We do not observe any shift in these measurements which is an indication of a weak resonance inelastic scattering process

## CONCLUSION

A reliable R matrix analysis of the experimental results [1] could not be fulfilled because of unknown influence of the higher excited states. The present measurements, performed at a higher energy of the $^{17}$O beam, showed an observation of very strong resonances at high energies of 8.6 MeV. At this energy, the density of states in $^{21}$Ne is well over 100 levels per 1 MeV. These resonances are also over 10 MeV above the neutron decay threshold. Therefore, new data are an indication for a remarkable manifestation of $α$-cluster structure in $^{21}$Ne. The data will be analyzed in the framework of R-matrix approach.

## ACKNOWLEDGMENTS

The authors gratefully acknowledge the support and help during experiment by the staff of INFN-LNS. This material is based upon work supported by the Ministry of Education and Science of the Republic of Kazakhstan state-targeted program №BR05236454, Nazarbayev University ORAU project "Pulsed ion beam neutralization by volumetric plasma" and grants of the authorized representative of the government of the Republic of Kazakhstan at Joint Institute for Nuclear Research.